# Beating Temporal Phase Sensitivity Limit in Off-axis Interferometry based Quantitative Phase Microscopy

Yujie Nie,[1,2] Renjie Zhou,[1,2,a]

[1]Department of Biomedical Engineering, The Chinese University of Hong Kong, Shatin, New Territories, Hong Kong SAR, China

[2]Shun Hing Institute of Advanced Engineering, The Chinese University of Hong Kong, Shatin, New Territories, Hong Kong SAR, China

[a]Author to whom correspondence should be addressed: rjzhou@cuhk.edu.hk

**Abstract**

Phase sensitivity determines the lowest optical path length (OPL) value that can be detected from the noise floor in a quantitative phase microscopy (QPM) system. The temporal phase sensitivity is known to be limited by both photon shot-noise and a variety of noise sources from electronic devices and environment. To beat temporal phase sensitivity limit, we explore different ways to reduce different noise factors in off-axis interferometry-based QPM using laser-illumination. Using a high electron-well-capacity camera, we measured the temporal phase sensitivity values using non-common-path and common-path interferometry based QPM systems under different environmental conditions. A frame summing method and a spatiotemporal filtering method are further used to reduce the noise contributions, thus enabling us to push the overall temporal phase sensitivity to less than 2 picometers.

## I. Introduction

Measuring morphological changes in time is important for studying cellular activities and material processes. Such changes can be quantified through quantitative phase microscopy (QPM). As a label-free imaging method, QPM precisely maps the optical path length (OPL) through interferometry or computation from intensity measurements [1-4]. However, measuring intrinsically weak nanoscale changes (e.g. protein aggregation [5], virus particle dynamics [6] and spiking-induced membrane fluctuations [7]) without exogenous labelling is very challenging with conventional QPM methods due to the low signal-to-noise-ratio (SNR). Many factors can contribute to noise during phase measurements, such as photon shot-noise, electronic source (e.g., device instability, camera noise, and quantization errors), light source instability, coherence properties, and other environmental factors (e.g., mechanical vibrations and air-density variations) [8-12].

Realizing the influence of different noise factors, over the past 15 years many active efforts have been made to enhance the phase sensitivity in off-axis interferometry based QPM systems (off-axis QPM in short), especially the temporal phase sensitivity. S. Chen *et al*. [13] derived the best achievable phase sensitivity using the Fourier transform-based algorithm in off-axis digital holography. N. T. Shaked *et al.* [14] reported placing the interferometer in a vacuum-sealed enclosure to avoid the influence of the air flow and used a floating optical table to damp the device oscillations to some extent. R. Zhou *et al*. [15] explored the use of high stability laser sources (i.e., temperature and current controlled semiconductor lasers) and broadband light sources to minimize the phase noise. G. Popescu *et al.* [16] developed diffraction phase microscopy (DPM) with a common-path off-axis interferometry geometry to diminish the mechanical vibrations. By introducing white light illumination sources, both temporal and spatial sensitivities have been significantly improved in DPM due to the reduction of speckle noise inherent to laser sources [9]. H. Majeed *et al*. [17] demonstrated that applying spatiotemporal filtering can push temporal phase sensitivity to 5 picometers. Despite the efforts on using low noise cameras, high stability light sources, white-light sources, common-path interferometry geometries, and digital filtering schemes to reduce the phase noise, the photon shot noise plays a dominant role in limiting phase sensitivity [9]. On the other hand, the noise from environmental factors are often difficult to control. P. Hosseini *et al.* [18] reported that using a high electron-well-capacity camera in a common-path off-axis QPM system results in much higher temporal sensitivity than using a normal camera. T. Ling *et al*. [19]

achieved an even higher phase sensitivity value of less than 4 picometers by applying a spike-triggered averaging method with an ultra-fast speed camera. To observe the roadmap in improving the temporal phase sensitivity in off-axis QPM, we highlight several representative works and extracted their reported temporal phase sensitivity in OPL values (Fig. 1). To alternatively characterize phase sensitivity, we define a new phase SNR metric as $SNR = 10log_{10}(2\pi/\varphi) = 10log_{10}(\lambda/OPL)$, where $\varphi$ is the phase noise value in radians, OPL is related to $\varphi$ through $OPL = \varphi \cdot \lambda/2\pi$, and $\lambda$ is the central wavelength of the light source in free space. Note that in [18, 19], where the phase SNR values are calculated to be over 45 dB, the systems were claimed operating under the photon shot-noise limit.

Based on previous experimental observations and theories, in this work we mainly explore how mechanical vibrations from different environmental sources and photon shot-noise limit the temporal phase sensitivity in off-axis QPM with laser-illumination. We also propose and demonstrate several methods or in combination to push the temporal phase sensitivity limit in this type of system. The temporal phase sensitivity limit is first explored and compared in common-path (i.e., DPM configuration) and non-common-path interferometry based off-axis QPM systems when operating under different environmental conditions. A spatiotemporal filtering method and a frame summing method are then used to further push the temporal phase sensitivity to around 2 picometers, which to our best knowledge is better than any reported results for off-axis QPM.

## II. Phase Sensitivity Analysis and Comparison between Common-path and Non-common-path Off-axis QPM Methods

To first explore how the mechanical vibration from environment affects the phase sensitivity, we constructed two different QPM systems as schematically shown in Fig. 2 (a)-(c). Fig. 2(a) and (b) are configured to form a QPM design with a common-path Mach-Zehnder interferometer geometry (i.e., a DPM system), while Fig. 2(a) and (c) are configured to form a QPM design with a regular non-common-path Mach-Zehnder interferometer geometry. In both QPM systems, a single-mode fiber coupled laser with a central wavelength of 532 nm is used as the illumination source (MGL-FN-532, CNI Laser). After the laser beam is collimated by lens L2, a 4f system, consisting of FL1 and object lens OL1 (EC Plan-Neofluar 10×/0.3 Ph1 M27, Zeiss), is used to bring the beam to the sample plane with a relatively uniform intensity distribution. The scattered light from the sample is collected by an imaging objective OL2 (EC Plan-Neofluar 40×/0.75 M27, Zeiss) and an intermediate image is formed after the tube lens TL1 (TL1 and OL2 forms a 4f system). From the intermediate image plane, an off-axis interferometer is constructed for retrieving the image field phase. For the common-path design (Fig. 2(b)), a diffraction grating (DG) is placed precisely at the intermediate image plane to divide the image field into multiple orders with each order containing the same image information (note that only the $0^{th}$ order and $1^{st}$ order beams are used later). At the Fourier plane after L4, a 10 μm diameter pinhole (PH) is used as a low-pass filter to convert the $0^{th}$ order beam into a reference beam that does not contain sample information, while the $1^{st}$ order beam remains unfiltered. Lens L4 and L5 form another 4f system to relay the sample to the final image plane, where interferograms are finally formed on a camera. For the non-common-path design (Fig. 2(c)), the illumination beam is separated into two by a 1×2 fiber coupler. The reference beam is collimated by lens L3 and then combined with the sample beam through a beam splitter (BS). P1 and P2 are polarizers for adjusting the intensities of the two beams to achieve the best fringe contrast. In this study, a high electron-well-capacity camera with a full-well-capacity of 2 million electrons (Q-2A750/CXP, Adimec) is used to capture interferograms in both QPM systems.

To quantify temporal phase sensitivity and spatial phase sensitivity of both QPM systems, we measured sample-free interferograms under four different environmental conditions: (1) during day & air conditioner on; (2) day & air conditioner off; (3) night & air conditioner on; (4) night & air conditioner off. Under each condition, 25 sample-free interferogram stacks were recorded at 500 frames per second (fps) with an exposure time of 1337 μs. Each stack contains 601 interferograms and the image size is 1024×1024 pixels. In each stack, the first frame is used for calibrating the phase map retrieved using a Fourier transform based algorithm [20]. After obtaining the phase frames in each stack, we computed the frequency spectrum of the mean phase values of each phase map. The averaged frequency spectra of 25 stacks are shown in

Fig.2. (d) and (e) for common-path and non-common-path systems, respectively. We found the environmental disturbance (mainly the mechanical vibrations of the building structure) is prominent at the 15-30 Hz frequency band. In both the low frequency region and the high frequency region, spectrum peaks are observed in the non-common-path system. From the spectrum analysis, we conclude that common-path design can achieve better phase sensitivity through isolating both high frequency and low frequency noise induced by mechanical vibrations from the environment.

To quantify the phase noise under each experimental condition, we calculated the phase sensitivity values in terms of OPL values and summarized them in Table 1 ($OPL_t$: temporal phase sensitivity; $OPL_s$: spatial phase sensitivity). At condition 4 (night & air conditioner off) when environmental disturbance is minimized, the lowest phase noise occurred, and the best temporal phase sensitivity was around 0.08 nm in average. The results also indicate that non-common-path system is more sensitive to environmental disturbance as expected. This specifically designed experiment serves as a confirmation of common-path QPM design in isolating environmental vibrations. When environment factors are minimized, the temporal phase sensitivity is mainly determined by the photon shot-noise which originates from the discrete nature of photons. In off-axis QPM, the temporal phase sensitivity is related to the effective electron-well-capacity ($N_{eff}$) of the camera as $\varphi \approx 1/\sqrt{N_{eff}}$ [18]. $N_{eff}$ can be derived from the histogram of a typical interferogram as shown in Fig. 3(b). For our common-path system, we estimate $N_{eff} = |N_2 - N_1| \approx 500000$ electrons. Note that in off-axis QPM, the phase is interpreted from fringes and the phase value at each pixel is affected by the whole diffraction spot area. Therefore, the phase sensitivity needs to be weighted over all the pixels within one diffraction spot, i.e., $\varphi \approx 1/\sqrt{m \cdot N_{eff}}$, where m is the effective number of pixels [21]. In our QPM system, there are around 4.3 pixels for each fringe period and 12 pixels projected in each diffraction-limited spot, we determined m to be 14. Then the theoretical temporal phase sensitivity value, $OPL_t$ is calculated to be around 0.032 nm for our system. From the values in Table 1, we found that best matched result is from the common-path system that has minimum noise contribution from the environmental vibrations, showing $OPL_t$ of 0.044 nm that is close to the photon shot-noise limit.

### III. Methods for Phase Sensitivity Enhancement

#### A. Frame Summing Method

To improve the phase sensitivity in common-path QPM, we explore a frame summing method that is effectively increasing the electron-well-capacity. The procedure of this method is illustrated in Fig. 3. (a). We initially divide the raw interferograms into several groups and sum the images to generate a summed interferogram. Then we use the first summed interferogram as the reference (indicated by the blue dotted box) for calibrating the phase images. The phase sensitivity values are then calculated based on the phase image stack. A stack of 600 sample-free interferograms were acquired at 500 fps with an exposure time of 1337 μs from the DPM system. As shown in Table 2, more than 4 times improvement in $OPL_t$ can be achieved to around 8 picometers by summing 100 frames. The result is smaller than the theoretical improvement of 10 times [10], which is likely due to the residual environmental vibration induced noise. Note that as temporal phase sensitivity improves, spatial phase sensitivity also improves.

#### B. Spatiotemporal Filtering Method

To further beat the temporal phase sensitivity limit, we applied a temporal and spatial bandwidth (BW) filtering method (or spatiotemporal filtering method in short [17]) to diminish the vibrational noise. To determine how the spatial and temporal scales affect the sensitivity, we draw a 3D spatiotemporal map by taking the Fourier transform along $x$, $y$, and $t$ of the selected image stack as shown in Fig. 4(a). The spatial bands selected are circles centered at the origin with radius of $2\pi$, $\pi$, and $\pi/2$. When selecting the low frequency bandwidth (15-55 Hz band with noticeable vibration-induced noise according to Fig. 2(d)), the $OPL_t$ values are reduced to 5.6 picometers ($2\pi$), 4.4 picometers ($\pi$), and 3.4 picometers ($\pi/2$) as we reduce the spatial BW. This indicates that the temporal phase sensitivity can be improved through sacrificing spatial resolution for a selected temporal frequency range. When selecting the high temporal frequency bandwidth (180-220 Hz band with negligible vibration-induced noise), the best $OPL_t$ value is 1.9 picometers when the spatial band is $\pi/2$. This analysis indicates that if there is flexibility in selecting the spatial and temporal frequency band, one can beat the phase sensitivity limit.

## IV. Implementation of Sensitivity Enhancement Methods for Measuring Cell Dynamics

To demonstrate the feasibility of the proposed methods for imaging biological specimens, we measure the membrane displacement of live human red blood cells (RBCs) suspended in a phosphate-buffered saline solution (Fig. 5(a)). For the selected RBC (Fig. 5(b)), the diameter was approximately 7.8 μm and the average heights of the rim and the dimple areas were 2.15 μm and 1.36 μm, respectively. The morphological parameters are consistent with the literature [22]. We recorded 500 sample frames at 500 fps with 1337 μs exposure time and obtained the OPL displacement map. We selected 8 representative regions with 30×30 pixels for comparing the phase sensitivity improvement. A higher displacement is observed at the rim of the RBC than in the dimple region. The original $OPL_t$ values of region 1 and 2 are 3.06 nm and 2.88 nm. By dividing the refractive index difference of 0.063 between the RBC and medium, we obtain a membrane displacements of 48.6 nm and 45.7 nm that are comparable to the values in literature [23]. Referring to [24], we selected a 0-25 Hz temporal BW and a $3\pi$ radius spatial BW central circle for spatiotemporal filtering. A 100-frame summing method is then applied to the reconstructed phase stack. At the background region 7 and 8, more than 3 times improvement in temporal phase sensitivity has been achieved to 0.13 nm and 0.15 nm, respectively. The sensitivity enhancement is not close to the sample-free theoretical limit, which might be due to the following two reasons: (1) we did not eliminate the frequency at around 20 Hz where there is vibration-induced noise; and (2) the liquid medium and sample motions prevented us to achieve temporal phase sensitivity close to the theoretical limit (a follow up study is being pursued by us).

## V. Discussion

In this letter, we have quantified temporal phase sensitivity of both common-path and non-common-path off-axis QPM systems under different environmental conditions, demonstrating that the non-common-path system is more sensitive to environmental disturbance. We applied the frame summing method and spatiotemporal filtering method to push the temporal phase sensitivity to a certain extent, while compromising the temporal and spatial resolution. Although we have demonstrated a potential to achieve less than 2 picometers of temporal phase sensitivity, there are still issues to be solved when applying our method for nanoscale dynamics observation in practical applications. Note that part of this work was presented at SPIE Photonics West 2020 and published with a different format in a conference proceeding [21].

## Acknowledgements

This work was supported by Croucher Foundation (CM/CT/CF/CIA/0688/19ay), Shun Hing Institute of Advanced Engineering (BME-p3-18), and Innovation and Technology Commission - Hong Kong (ITS/098/18FP, ITS/394/17).

## Data Availability Statements

The data that support the findings of this study are available from the corresponding author upon reasonable request.

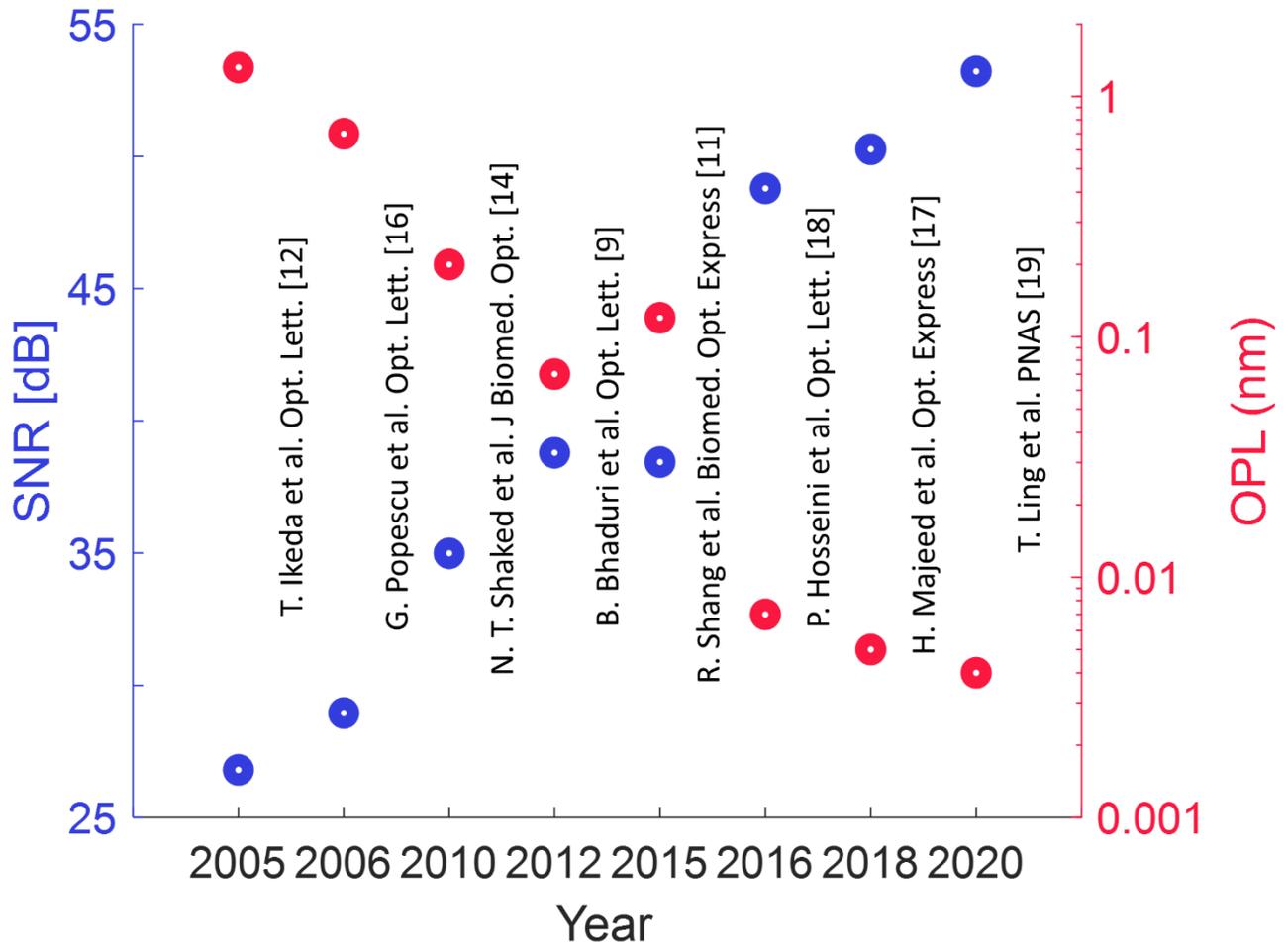

Fig. 1. Selected literature showing the roadmap of temporal phase sensitivity improvement in off-axis QPM methods.

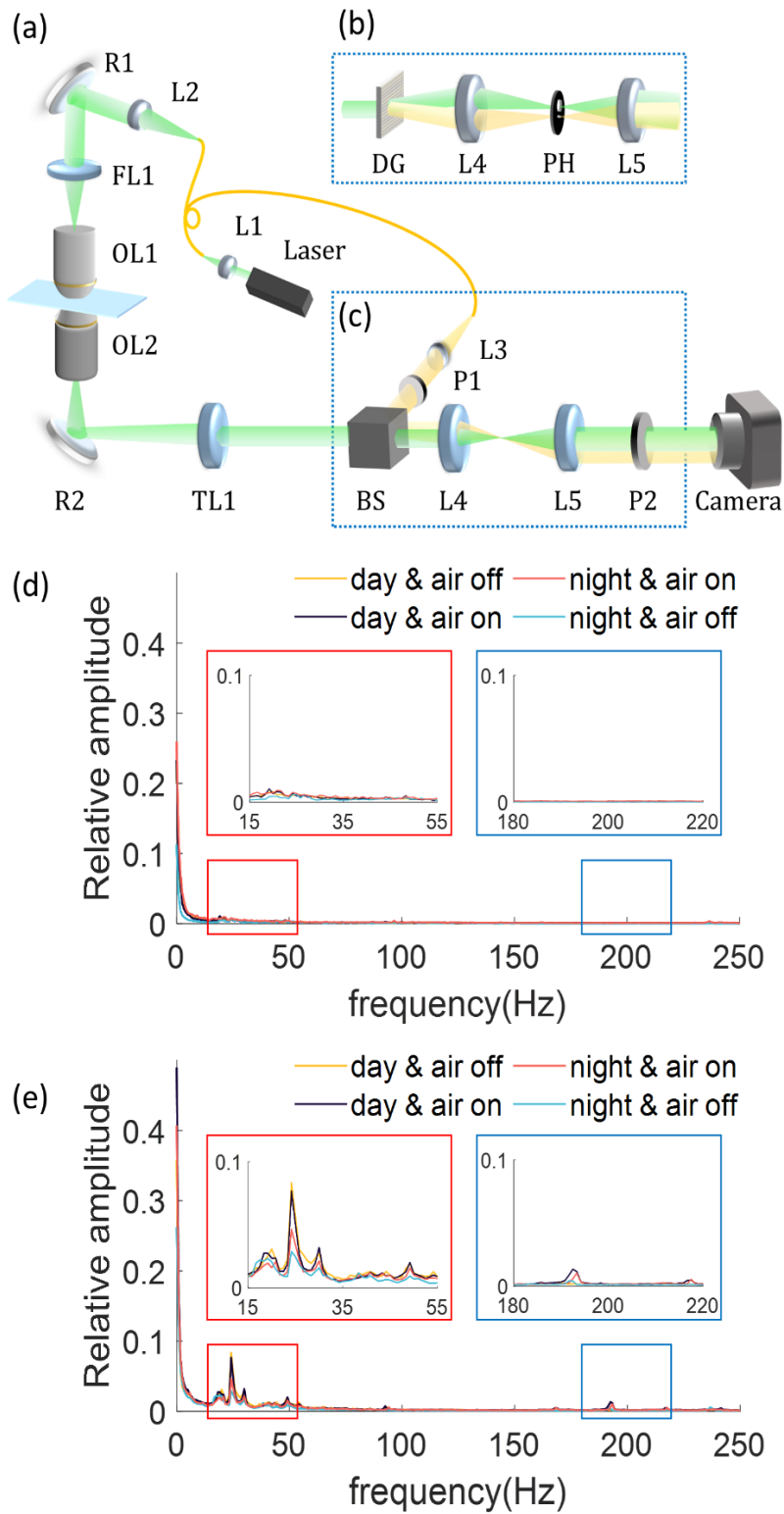

Fig. 2. (a)-(c) Schematic design of the microscopy system, common-path interferometer, and non-common-path interferometer, respectively. (d) (e) Averaged frequency spectra of the phase maps measured with common-path and non-common-path QPM systems, respectively. L: lens; R: reflector; FL: field lens; OL: objective lens; TL: tube lens; BS: beam splitter; P: linear polarizer; DG: diffraction grating; PH: pinhole.

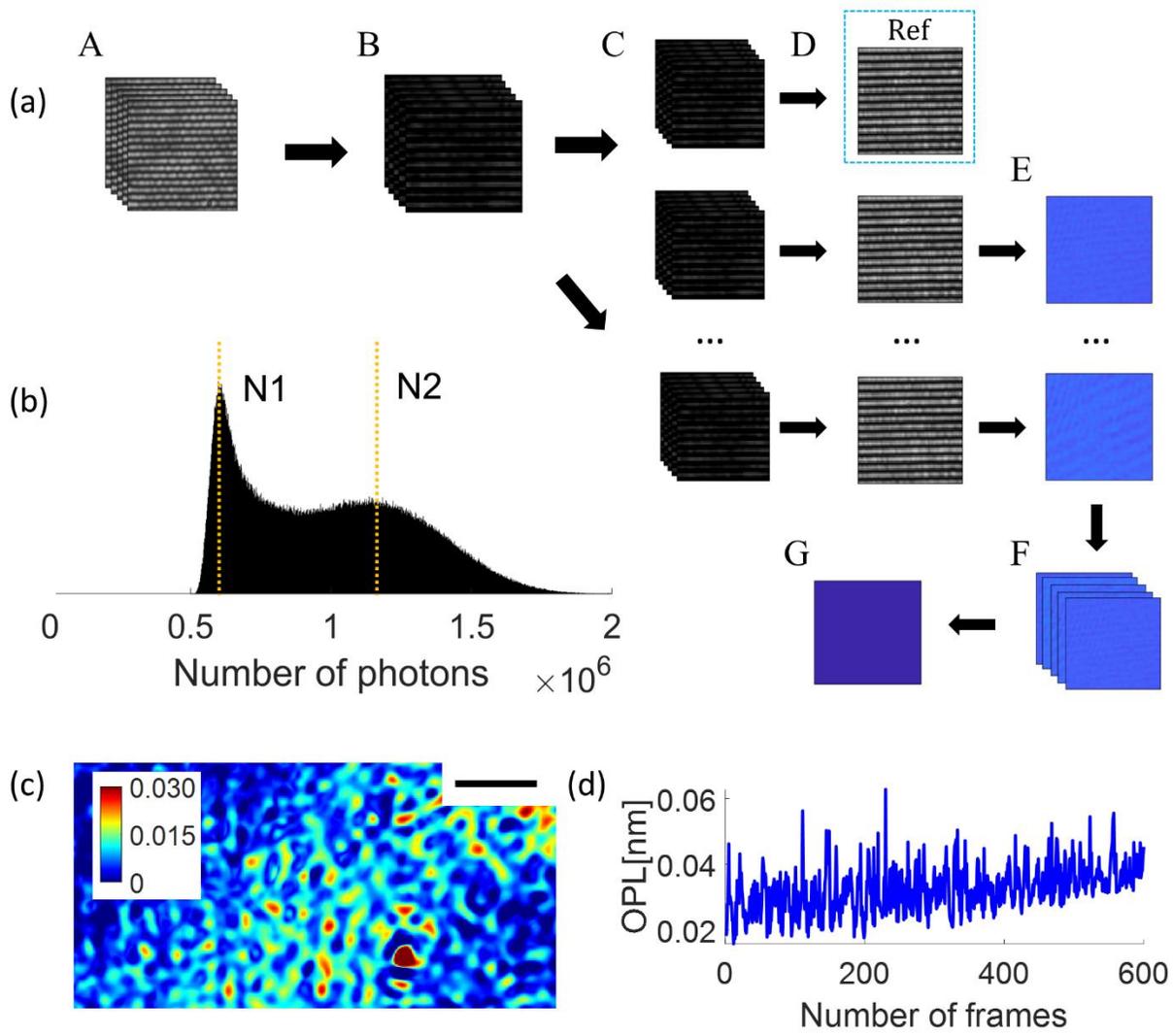

Fig. 3. (a) Flow chart of the frame summing algorithm. A. n×T raw interferograms. B. Normalized raw interferograms by subtracting frame mean value. C. Divide B into n groups with each group containing T frames. D. Summed interferograms of each group in C. E. Use the first summed interferogram for phase calibration and obtain n-1 phase maps. F. Normalized phase stack by subtracting the frame mean value. G. Standard deviation map of F. (b) Histogram of the intensity distribution of a raw interferogram. (c) A 500×250 pixels sample-free OPL standard deviation map in unit of nm. Scale bar: 5 μm. (d) Variations in the frame mean OPL over 600 frames.

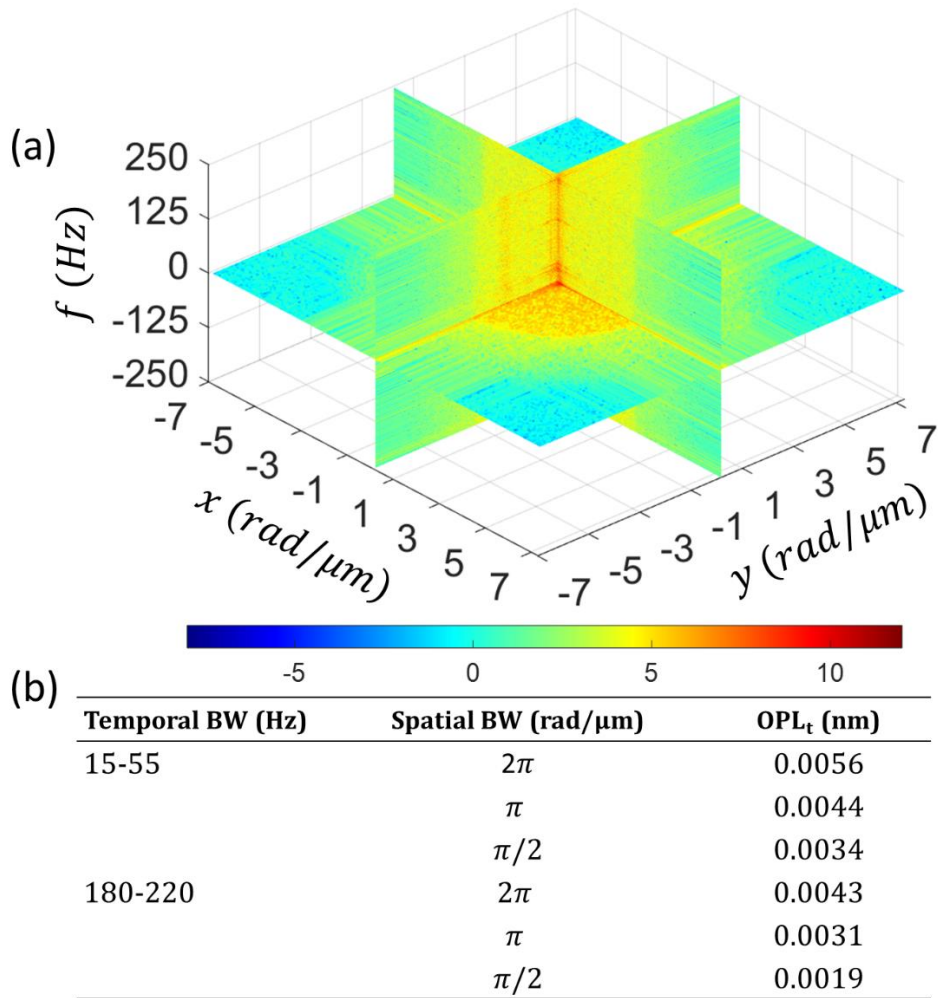

Fig. 4. (a) Spatiotemporal spectrum along three different planes in 3d frequency domain. Colormap is in log scale. (b) Bandpass filtering over the selected spatiotemporal bands. BW: bandwidth.

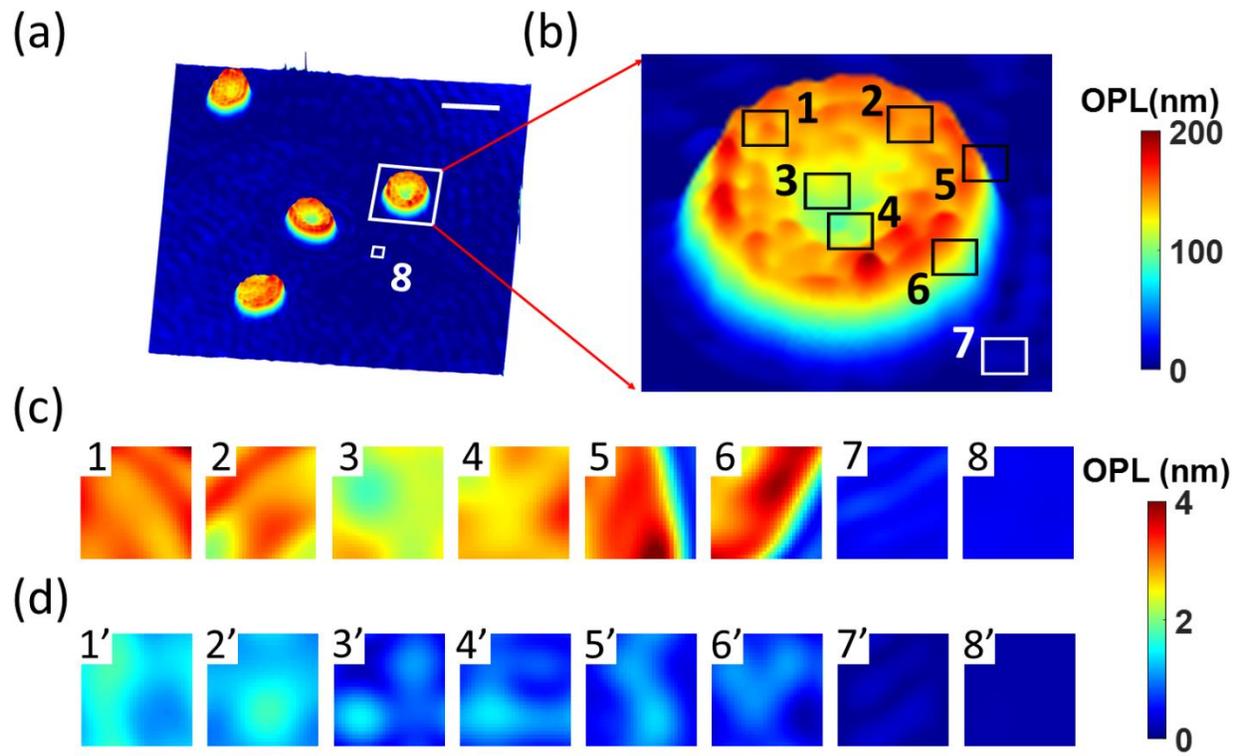

Fig. 5. (a) Phase map of human RBCs. Scale bar: 10 μm. (b) A selected region from (a). (c) (d) Temporal phase map of region 1-8 as indicated in (a)(b) without and with phase sensitivity enhancement processing, respectively.

Table 1. Phase sensitivity in different environmental conditions

|  | Non-common-path design | | Common-path design | |
| --- | --- | --- | --- | --- |
|  | $OPL_t$ (nm) | $OPL_s$ (nm) | $OPL_t$ (nm) | $OPL_s$ (nm) |
| Day & Air on | 0.25±0.08 | 0.49±0.23 | 0.15±0.08 | 0.22±0.10 |
| Day & Air off | 0.21±0.06 | 0.38±0.20 | 0.11±0.05 | 0.21±0.14 |
| Night & Air on | 0.22±0.07 | 0.38±0.20 | 0.12±0.04 | 0.23±0.12 |
| Night & Air off | 0.11±0.03 | 0.29±0.17 | 0.08±0.03 | 0.15±0.09 |

Table 2. Phase sensitivity improved through frame summing

|  | $OPL_t$ (nm) | $OPL_s$ (nm) |
| --- | --- | --- |
| 1 frame | 0.0353 | 0.0516 |
| 20 frames | 0.0120 | 0.0236 |
| 50 frames | 0.0099 | 0.0206 |
| 100 frames | 0.0083 | 0.0197 |